%% file: 96zr_pub_v12.tex
\documentclass[final,3p,times]{elsarticle}%
\usepackage{graphicx}%
\usepackage{subfigure}%
\usepackage{amssymb}%
\usepackage{amsmath}%
\usepackage{hyperref}%
\usepackage{booktabs}
\input{newcommands}%
\journal{Nuclear Physics A}%
\begin{document}%
\begin{frontmatter}%
\title{Measurement of the two neutrino double beta decay half-life of Zr-96 with the NEMO-3 detector}%
\input{nemo3-authors}%
\begin{abstract}%
Using 9.4\,g of \zr\ isotope and 1221 days of data from the NEMO-3 detector corresponding to 0.031\,\kgy, the obtained \tn\ decay half-life measurement is \ttn\ = \zrhalf. Different characteristics of the final state electrons have been studied, such as the energy sum, individual electron energy, and angular distribution. The $2\nu$ nuclear matrix element is extracted using the measured \tn\ half-life and is \mtn\ = \zrnme. 
Constraints on $0\nu\beta\beta$ decay have also been set.
\end{abstract}%
\end{frontmatter}%

\section{Introduction}
The recent observation of neutrino flavor oscillations and the resulting measurements of the neutrino mass squared differences~\cite{ Fogli:2005cq} have motivated renewed experimental efforts to measure the absolute neutrino mass. The fundamental Dirac or Majorana~\cite{ Majorana:1937vz} nature of the neutrino also remains indeterminate. Neutrinoless double beta decay ($0\nu\beta\beta$) is the only practical means of determining the nature of the neutrino and one of the most sensitive probes of its absolute mass in the case of Majorana neutrinos. The mechanism in which a light Majorana neutrino is exchanged~\cite{ Furry:1939qr} is most commonly discussed and the half-life in this case is given by
\begin{equation}
[T_{1/2}^{0\nu}]^{-1} = G^{0\nu} \vert M^{0\nu} \vert ^{2} \langle m_{\beta\beta} \rangle^{2}\,,
\label{equ:0vbb} \end{equation}
where $G^{0\nu}$ is the precisely calculable phase-space factor (proportional to $Q_{\beta\beta}^{5}$), $M^{0\nu}$ is the nuclear matrix element (NME) and $\langle m_{\beta\beta} \rangle$ is the effective Majorana mass of the electron neutrino. Other possible mechanisms for $0\nu\beta\beta$ include, for example, right-handed currents, Majoron emission and R-parity violating supersymmetry. In all mechanisms, the $0\nu\beta\beta$ process violates lepton number and is a direct probe for physics beyond the Standard Model.
%

Measurement of the $2\nu\beta\beta$ process is important because it is an irreducible background component to $0\nu$ mechanisms. Double beta decay ($2\nu\beta\beta$) allows experimental determination of the NME for this process ($M^{2\nu}$), which leads to the development of theoretical schemes for NME calculations for $2\nu\beta\beta$ and $0\nu\beta\beta$~\cite{ Kortelainen:2007mn, Simkovic:2007vu, Rodin:2007fz}. The precision with which the lepton number violating parameter, such as $\langle m_{\beta\beta} \rangle$, can be measured depends crucially on knowledge of $M^{0\nu}$. Presented here are the results of observations of $^{96}$Zr obtained with the NEMO-3 tracking plus calorimeter detector.

\section{NEMO-3 Experimental Apparatus}
A detailed description of the NEMO-3 detector and its performance can be found in~\cite{ Arnold:2004xq}, while the most salient properties are mentioned here. The detector is located in the Modane Underground Laboratory (LSM) 4800 meters water equivalent below ground and has been acquiring data since February 2003. It is a cylindrical detector ($\varnothing~5 \times 2.5$\,m) holding 10\,kg of enriched isotopes. The tracking volume contains $\sim$\,6000 drift cells operating in Geiger mode (Geiger cells) enclosed by $\sim$\,2000 polystyrene scintillator blocks making up the calorimeter. The detector is enclosed in a solenoid which generates a 25\,Gauss magnetic field parallel to the Geiger cells. The transverse and longitudinal resolution of the tracker is 0.6\,mm and 0.3\,cm ($\sigma$) respectively. The calorimeter energy resolution and timing resolution is 14--17\% (FWHM at 1\,MeV) and 250\,ps ($\sigma$ at 1\,MeV) respectively.

The majority of the $\beta\beta$ isotope mass is $^{100}$Mo but other isotopes include $^{82}$Se, $^{116}$Cd, $^{130}$Te, $^{150}$Nd, $^{96}$Zr, and $^{48}$Ca. The experimental signature of $0\nu\beta\beta$ is two electrons with the energy sum equaling the $Q_{\beta\beta}$ of the decay. $^{96}$Zr is of particular interest due to its high $Q_{\beta\beta} = 3350.0 \pm 3.5$\,keV which is greater than the decay energies of most contributing background sources, and the large phase-space factor which is proportional to $Q_{\beta\beta}^{5}$. The total mass of the enriched ZrO$_{2}$ is 22.0\,g of which $9.4 \pm 0.2$\,g is $^{96}$Zr~\cite{ Arnold:2004xq}. NEMO-3 results thus far are published in~\cite{ Arnold:2005rz, Arnold:2006sd, Arnold:2006fk, Argyriades:2008pr}.

\section{Event Topology and Particle Identification}
The NEMO-3 detector is capable of sophisticated particle identification and event topology reconstruction. Electrons and positrons produce signals in both the calorimeter and Geiger cell tracker, while photons only create a signal in the calorimeter. Due to the 25 Gauss magnetic field permeating the detector volume, the electron and positron discrimination efficiency is 97\% at 1\,MeV. Alpha particles ($\alpha$) are identified by the short distance ($\sim$\,20\,cm) they travel before quenching in the gas volume of the Geiger cells. Crossing electrons (an electron crossing the whole tracker volume and source foil to mimic a $\beta\beta$ event) are identified by the time-of-flight information from the two signaled calorimeter blocks. The event topologies studied in this analysis include the single electron channel ($1e$), the electron plus gamma channel ($e\gamma$), and the two electron channel ($ee$).

\section{Backgrounds in NEMO-3}
Studies have been carried out to identify the activities of the contributing backgrounds to NEMO-3~\cite{ Argyriades:2009vq}. The backgrounds are categorized as ``internal" or ``external." Internal backgrounds include isotopes decaying from within the source foil mimicking a $\beta\beta$ decay via M{\o}ller scattering, $\beta$ decay with internal conversion, or $\beta$ decay with Compton scattering of the de-excitation photon. Each of the seven $\beta\beta$ isotopes being measured at NEMO-3 has specific dominant internal backgrounds. External backgrounds include all decays originating from outside the source foil but still mimic a $\beta\beta$ event signature via double Compton scattering, Compton plus M{\o}ller scattering, or pair production. Charge identification via track curvature in the magnetic field tags pair production events. The two most detrimental contributers are $^{214}$Bi and $^{208}$Tl with respective $Q_{\beta}$ values of 3.27\,MeV and 4.99\,MeV. NEMO-3 component and source foil activities were measured with a high purity germanium detector (HPGe) and were subject to a selection process to optimize radio-purity.

\subsection{Radon ($^{222}$Rn)}
The first data acquisition period (Feb 2003 -- Oct 2004) is referred to as Phase-I and had a relatively high level of radon in the tracking volume with a total activity of 1200\,mBq. Radon ($^{222}$Rn) is particularly disruptive because it is a noble gas and its half-life of 3.82 days provides enough time to be outgassed from the surrounding rock and permeate the detector volume. Supporting evidence suggests~\cite{ Pagelkopf:2003ae} that a large fraction (87\%) of $\alpha$ decay daughters are positively charged and are attracted to electrically negative and grounded surfaces. NEMO-3 data are consistent with the radon daughters being deposited on the surfaces of reflecting wrapping around the scintillators, the drift cell cathode wires and the source foils~\cite{ Argyriades:2009vq}.

The second data acquisition period (Nov 2004 -- Dec 2007) is referred to as Phase-II and began with the installation of a radon purification facility to inject a flow of pure air around the detector. The purification facility suppresses the radon concentration in the immediate proximity of the detector by a factor of $\sim$\,1000. However, the outgassing of detector components releasing radon due to their internal contamination with the $^{238}$U\,--\,$^{226}$Ra chain leads to a smaller reduction factor inside the detector. The radon activity in the tracker volume decreased from 1.2\,Bq in Phase-I to 0.2\,Bq in Phase-II.

\section{Data Analysis}
All background and signal events are simulated with DECAY0~\cite{ Ponkratenko:2000um} which accurately reproduces energy and angular distributions of particles emitted in radioactive decays including $2\nu\beta\beta$ and theoretical $0\nu\beta\beta$ mechanisms. All generated particles are propagated through a full GEANT--3.21~\cite{ Brun:1987ma} description of the detector. The simulated Monte Carlo (MC) events are in the same format as the raw data from the NEMO-3 detector and both MC and real data are reconstructed with the same software package.

\subsection{Background Identification} \label{sec:bkgs}
One can measure the activities of the various background isotopes by the event topologies and kinematics determined by the selection criteria. All background isotopes are measured with the single electron ($1e$) and electron plus gamma ($e\gamma$) channels. A global analysis of the external background is discussed in~\cite{ Argyriades:2009vq}. $^{208}$Tl and $^{214}$Bi were independently measured using $e\gamma\gamma$, $e\gamma\gamma\gamma$, and $e\alpha$ channels. The so-called ``external background model" has been tested and validated using the dedicated sectors of ultra--pure Cu and Te foils in NEMO-3.

Limits have been placed on the internal background activities of $^{96}$Zr by a high purity germanium (HPGe) detector, but ultimately the internal background activities are measured with the NEMO-3 apparatus. Internal background activities are measured in the $1e$ and $e\gamma$ channels. The $1e$ selection criteria are the following: one negatively charged particle track with length greater than 50\,cm originating from the $^{96}$Zr source foil and terminating at a scintillator, and an energy deposit $>$\,500\,keV in the scintillator associated with the track. The $e\gamma$ selection criteria are the following: one negatively charged particle track with length greater than 50\,cm originating from the $^{96}$Zr source foil and terminating at a scintillator, an energy deposit $>$\,200\,keV in the scintillator associated with the track, an energy deposit $>$\,200\,keV in a separate scintillator with no associated track, the cosine of the angle between the electron and gamma must be $<$\,0.9\,, and the time-of-flight information must be consistent with the electron and gamma originating from the same point in the source foil.  In both the $1e$ and $e\gamma$ channels (for quality control of the reconstructed track) we require at least one triggered Geiger cell in first two layers closest to the source foil and less than 3 triggered Geiger cells that are not associated with the reconstructed track.

\begin{figure}[htp]
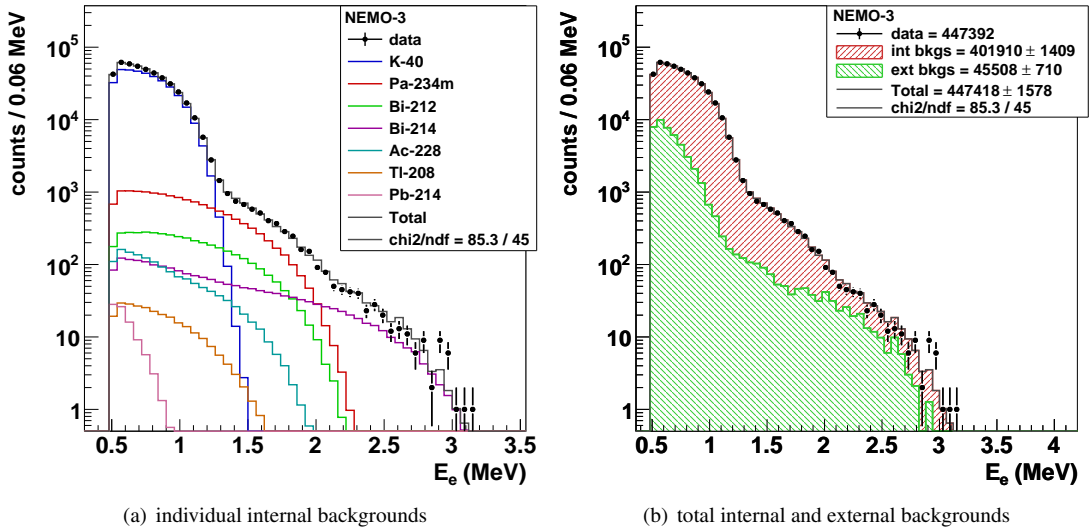
\centering
\subfigure[individual internal backgrounds]{%
	\includegraphics[width=\hw]{\eplots{Conf-internals}}}\hspace{0pc}
\subfigure[total internal and external backgrounds]{%
	\includegraphics[width=\hw]{\eplots{Conf-E-log-stack}}}
\caption{Energy spectra of the \zr\ backgrounds in the \e\ channel. Individual internal backgrounds are plotted (a) and the total background (b) is divided into 2 sub-groups of summed internal (int) and external (ext) components.} \label{fig:1e-channel} \end{figure}

\begin{figure}[ht!]
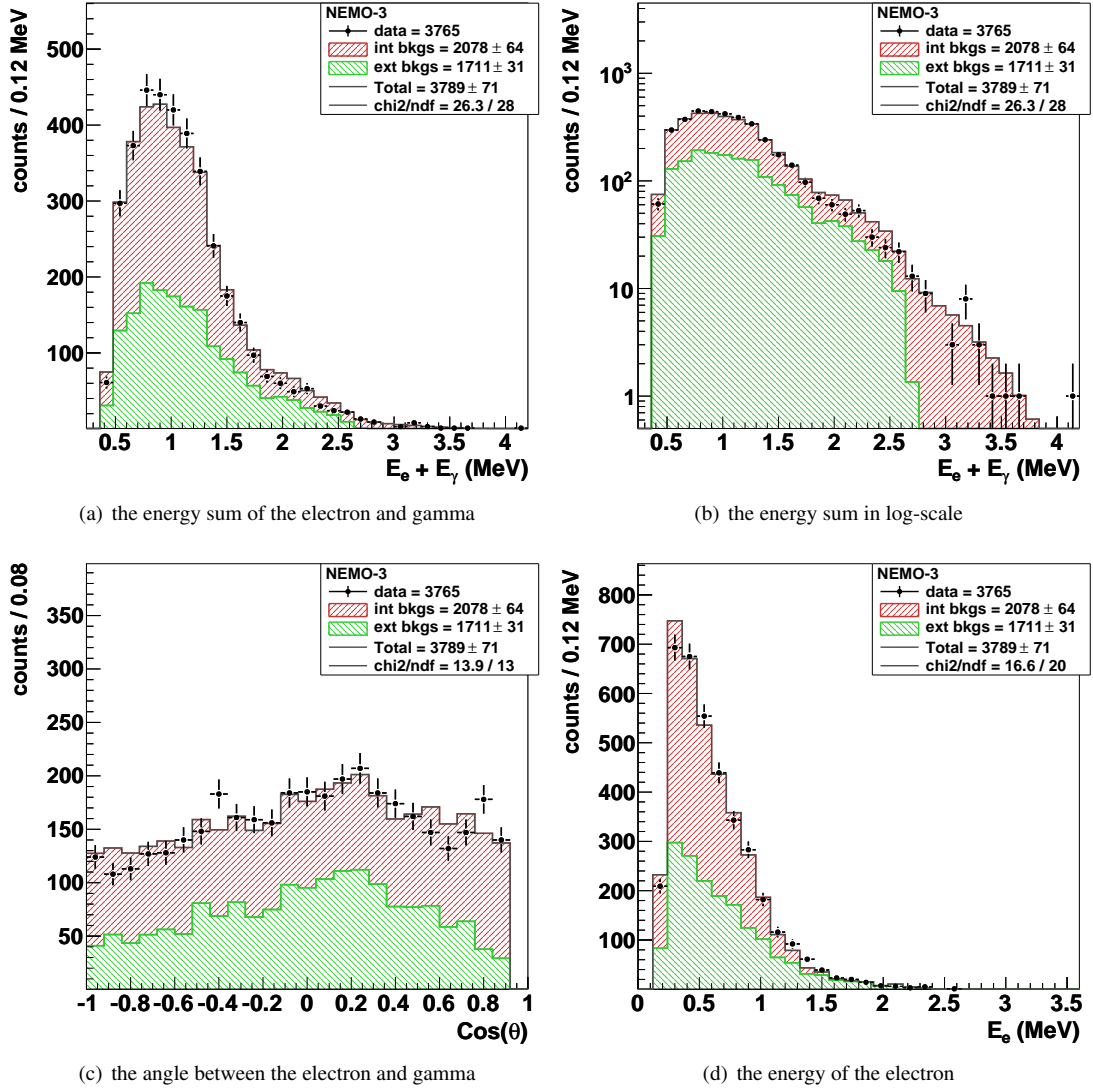
\centering
\subfigure[the energy sum of the electron and gamma]{%
	\includegraphics[width=\hw]{\egplots{Conf-totE-stack}}}\hspace{0pc}
\subfigure[the energy sum in log-scale]{%
	\includegraphics[width=\hw]{\egplots{Conf-totE-log-stack}}}\\
\subfigure[the angle between the electron and gamma]{%
	\includegraphics[width=\hw]{\egplots{Conf-cos-stack}}}\hspace{0pc}
\subfigure[the energy of the electron]{%
	\includegraphics[width=\hw]{\egplots{Conf-eleE-stack}}}
\caption{The $e\gamma$ channel displaying (a,b) the summed energy $E_{e} + E_{\gamma}$, (c) the angular distribution between the electron and gamma $\cos(\theta)$, and (d) the energy of the electron $E_{e}$. As in Figure~\ref{fig:1e-channel}, the background contributions are divided into 2 sub-groups of summed internal (int) and external (ext) components.} \label{fig:eg-channel} \end{figure}

The internal background activities are distinguished and measured due to contrasting $Q_{\beta}$ values and $1e$ and $e\gamma$ energy spectra of the isotopes. Equilibrium within a decay chain implies specific isotope activities to be correlated. $^{228}$Ac, $^{212}$Bi, and $^{208}$Tl are part of the $^{232}$Th chain and separated by short half-lives, therefore $^{228}$Ac and $^{212}$Bi activities are set equal and $^{208}$Tl is set to its branching ratio of 36\%. $^{214}$Bi and $^{214}$Pb belong to the $^{238}$U chain and are set equal. $^{234m}$Pa is also part of the $^{238}$U decay chain but equilibrium with $^{214}$Bi cannot be assumed due to the large half-life of the intermediate isotope $^{226}$Ra. Within this background model, contributions from the above isotopes to the $1e$ and $e\gamma$ channels have been fitted to experimental data over the entire energy region leaving the activities of the isotopes floating. Figure~\ref{fig:1e-channel} shows the goodness of fit of the $1e$ channel and has a $\chi^{2} = 85.3 / 45$. Figure~\ref{fig:eg-channel} shows the $e\gamma$ channel and has a $\chi^{2} = 26.3 / 28$. The individual and summed energy distributions of electrons and gammas as well as the angular distribution between them are plotted. The measurements of the internal $^{96}$Zr contamination obtained in the $1e$ and $e\gamma$ channels compared with previously obtained HPGe limits in Table~\ref{table:internals} provide a cross-check for the NEMO-3 measurements. The obtained numbers are in agreement with the $^{214}$Bi and $^{208}$Tl activities (0.17\,$\pm$\,0.05 and 0.08\,$\pm$\,0.01\,mBq respectively) reported in~\cite{ Argyriades:2009vq} where more restrictive energy intervals and different event topologies were used to identify signatures of the isotopes.
\begin{table}[htp]\centering
\caption{Internal contamination of the $^{96}$Zr foil measured with NEMO-3 in the $1e$ and $e\gamma$ channels under the assumptions of the background model described in~\ref{sec:bkgs}. Total source activities are given in milli-Becquerels (mBq) and the NEMO-3 measurements are compared to HPGe limits at 95\% confidence level.} \begin{tabular}{lcc} \\
\toprule
Isotope ~~~~~ & NEMO-3 (mBq)         & HPGe (mBq) \\
\midrule 
$^{228}$Ac    &  0.25\,$\pm$\,0.02    & $<$\,0.75  \\
$^{212}$Bi    &  0.25\,$\pm$\,0.02    & $<$\,0.75  \\
$^{208}$Tl    & 0.091\,$\pm$\,0.007   & $<$\,0.23  \\
$^{214}$Bi    &  0.19\,$\pm$\,0.02    & $<$\,0.45  \\
$^{214}$Pb    &  0.19\,$\pm$\,0.02    & $<$\,0.45  \\
$^{40}$K      &  19.7\,$\pm$\,0.1~    & $<$\,19    \\
$^{234m}$Pa   &  0.49\,$\pm$\,0.01    & $<$\,6.6   \\
\bottomrule
\end{tabular} 
\label{table:internals} \end{table}

The adjacent $\beta\beta$ source isotopes ($^{150}$Nd and $^{48}$Ca) and their associated internal backgrounds contribute events that pass the $^{96}$Zr selection criteria due to the positional resolution of the Geiger cell tracker and accuracy of the reconstructed event vertex. These events have been studied and contribute $\sim$\,1\% in the $1e$ channel and $\sim$\,7\% in the $e\gamma$ channel and are included in the background description for $^{96}$Zr.

\section{Results}
\subsection{Double Beta Decay of 96-Zr}
The selection criteria for $ee$ events are the following: two negatively charged particle tracks with lengths greater than 30\,cm, both tracks originating from the $^{96}$Zr foil and terminating at independent scintillators, energy deposits $> 200$\,keV in the scintillators associated with the tracks, each track has at least one triggered Geiger cell in first two layers closest to the source foil, and the time-of-flight information must be consistent with the two electrons originating from the same point on the source foil.

\begin{figure}[ht!]
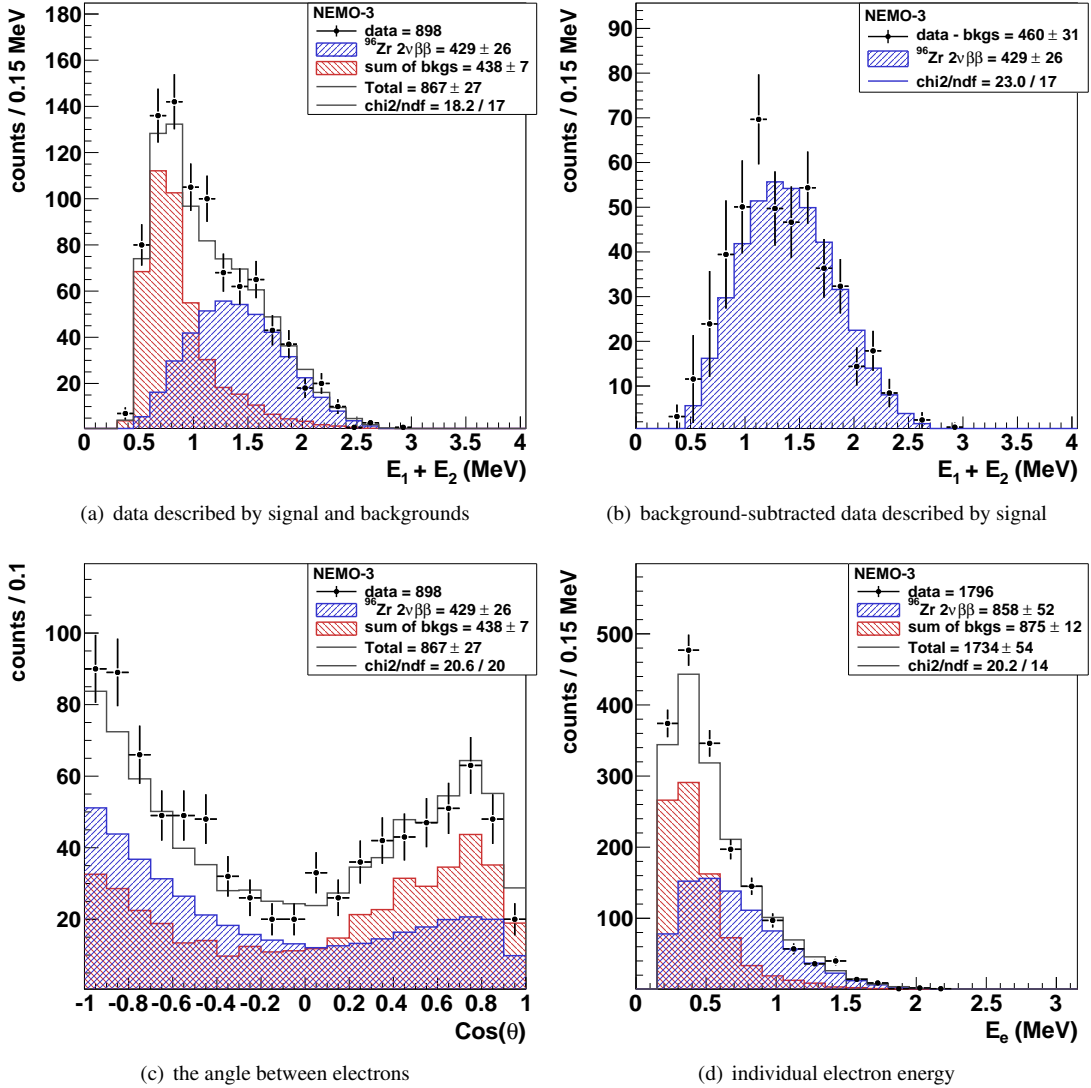
\centering
\subfigure[data described by signal and backgrounds]{%
	\includegraphics[width=\hw]{\eeplots{BKG-add-totE}}}\hspace{0pc}
\subfigure[background-subtracted data described by signal]{%
	\includegraphics[width=\hw]{\eeplots{BKG-sub-totE}}}\\
\subfigure[the angle between electrons]{%
	\includegraphics[width=\hw]{\eeplots{BKG-add-cos}}}\hspace{0pc}
\subfigure[individual electron energy]{%
	\includegraphics[width=\hw]{\eeplots{BKG-add-singE}}}
\caption{The (a) energy sum of both electrons $E_{1} + E_{2}$, (b) MC signal fit to background subtracted data, (c) angular distribution $\cos(\theta)$, and (d) individual electron energy $E_{e}$ for 1221 days of runtime in the $ee$ channel. The data are described by the sum of the expected backgrounds from MC and the $2\nu\beta\beta$ signal from the maximized log-likelihood fit.} \label{fig:ee-channel} \end{figure}

The distributions of the energy sum of the two electrons, energies of the individual electrons, and the angle between two electrons are shown in Figure~\ref{fig:ee-channel}. 898 data events have been selected after 1221 days of data taking with a total expected background of  $437.6 \pm 7.2$ events. A maximized binned log-likelihood fit to the energy sum spectrum is performed to estimate the $2\nu\beta\beta$ signal contribution. The likelihood fit predicts $429.2 \pm 26.2$ signal events (signal-to-background of 0.98) with a 7.5\% efficiency. The breakdown of individual background contributions is shown in Table~\ref{table:bkgs}.
\begin{table}[htp]\centering
\caption{The number of events expected for the $^{96}$Zr internal and external backgrounds in the $ee$ channel for 1221 days of runtime.}  \begin{tabular}{lrr} \\
\toprule
Background            & Expected $N_{bkg}$ & Eff. (\%)\\
\midrule
$^{228}$Ac            &  11.1\,$\pm$\,0.9  & 0.042 \\
$^{212}$Bi            &   9.6\,$\pm$\,0.7  & 0.036 \\
$^{208}$Tl            &   9.3\,$\pm$\,0.7  & 0.098 \\
$^{214}$Bi            &  22.8\,$\pm$\,2.5  &  0.12~~ \\
$^{214}$Pb            &   3.3\,$\pm$\,0.4  & 0.017 \\
$^{40}$K              & 280.0\,$\pm$\,2.4  & 0.014 \\
$^{234m}$Pa           &  38.3\,$\pm$\,0.7  & 0.074 \\
$^{48}$Ca internals   &   0.0\,$\pm$\,0.0  &  \\
$^{150}$Nd internals  &  37.6\,$\pm$\,3.2  &  \\
External              &  25.6\,$\pm$\,5.2  &  \\
\midrule
\textbf{Total}     & \textbf{437.6\,$\pm$\,7.2} &  \\
\bottomrule
\end{tabular} \label{table:bkgs} \end{table}

Limits on $0\nu$ processes have been obtained using a binned log-likelihood ratio (LLR) test statistics~\cite{ Junk:1999kv}. The results for \mbb, \rhc, and Majoron mechanisms are reported in~\tab{tab:big-fuckin-table}.
\begin{table}[htp]\centering
\caption{Summary of half-life limits $T_{1/2}$\,(yr.) evaluated at the 90\% CL for $0\nu\beta\beta$ mechanisms where $0^{+}_{gs}$(\mbb) is the standard $0\nu\beta\beta$ decay to the ground state, $0^{+}_{1}$(\mbb) is the first excited state, $0^{+}_{gs}$(\rhc) is the right-handed current decay to ground state and $2^{+}_{1}$(\rhc) is the first excited state. The spectral index ($n$) for the Majoron modes (\mnc) refers to the dependence of $G^{0\nu} \propto (Q_{\beta\beta} - T)^{n}$ where $T$ is the electrons' kinetic energy sum. The right-most column displays the previous best limit for comparison.} 
\begin{tabular}{lcc} \\
\toprule
\zn\ mechanisms & NEMO-3 Limit & Previous Limit \\
\midrule
$0^{+}_{gs}$ \mbb & $9.2 \times 10^{21}$ & $1.0 \times 10^{21}$\cite{ Arnold:1999vg} \\
$0^{+}_{1}$ \mbb & $2.2 \times 10^{20}$ & $6.8 \times 10^{19}$\cite{ Barabash:1994yh} \\
$0^{+}_{gs}$ \rhc & $5.1 \times 10^{21}$ & -- \\
$2^{+}_{1}$ \rhc & $9.1 \times 10^{20}$ & $3.9 \times 10^{20}$\cite{ Arnold:1999vg} \\
\midrule
Majoron modes & & \\
\midrule
$n = 1$ & $1.9 \times 10^{21}$ & $3.5 \times 10^{20}$\cite{ Arnold:2000uw} \\
$n = 2$ & $9.9 \times 10^{20}$ & -- \\
$n = 3$ & $5.8 \times 10^{20}$ & $6.3 \times 10^{19}$\cite{ Arnold:2000uw} \\
$n = 7$ & $1.1 \times 10^{20}$ & $5.1 \times 10^{19}$\cite{ Arnold:2000uw}  \\
\bottomrule
\end{tabular} \label{tab:big-fuckin-table} \end{table}
The limit on the $0\nu\beta\beta$ half-life is used to calculate an upper bound on the effective Majorana neutrino mass \mbb \lt 7.2 -- 19.5\,eV~\cite{ Kortelainen:2007mn, Simkovic:2007vu, Rodin:2007fz, Chaturvedi:2008zz} obtained with only 9.4\,g of source isotope.
%
%
%

\subsection{The Systematic Error}
The systematic error on the \tn\ measurement has been investigated. The main contribution is from the error on the tracking detector resolution and track reconstruction efficiency~\cite{Arnold:2004xq}. There is a 2\% uncertainty in the mass of \zr ~\cite{Arnold:2004xq}. The precision of the energy calibration of the calorimeter is 1\% and the effect was determined by coherently changing the gain of the PMTs \plm 1\% and observing the change in half-life. 

The systematic uncertainty of the external background model is considered. \bi{214}\ and \tl{208}\ in the tracking chamber show a discrepancy between the channels they are measured in. \bi{214}\ is measured in the \eg\ and \ea\ channels and the obtained values differ by \aprox 10\%~\cite{ Argyriades:2009vq}. \tl{208}\ is measured in the \egg\ and \eggg\ channels and the obtained values differ by \aprox 10\%~\cite{ Argyriades:2009vq}. A conservative estimation on the total external background uncertainty is therefore 10\% and is evaluated by fluctuating the external background \plm 10\%. The attributed uncertainty on the measured half-life is \plm 0.3\%. 

The systematic error on internal background model is estimated by the difference in measured activities in the \e\ and \eg\ channels. The difference never exceeds 5\% for the internal backgrounds, therefore the uncertainty on the \tn\ half-life is estimated by fluctuating the internal backgrounds \plm 5\% and recording the corresponding change in \tn\ half-life. 

The world's best \tn\ half-life measurements for \nd ~\cite{ FatemiGhomi:2009rs, Argyriades:2008pr} and \ca ~\cite{ shiva:2008} have been recently obtained. These isotopes neighbor the \zr\ source and are included as backgrounds. The uncertainty in their measured half-lives is applied and the change in the \zr\ half-life is noted. The \tn\ half-life of \nd\ is known to 10\% (including statistical and systematic errors) and contributes a \plm 0.7\% error on the obtained \zr\ half-life. The \tn\ half-life of \ca\ is known to 18\% (including statistical and systematic errors) and has a negligible contribution to the obtained \zr\ half-life.

\k\ is the dominant background in the \ee\ channel and a systematic effect is observed by changing the energy window of the likelihood fit to exclude energy sums below 1.1\,MeV. The strict energy window suppresses \k\ events and reduces the half-life dependence on the activity of \k. The obtained systematic uncertainties are listed in \tab{table:syst}\ and give a total systematic error of +\,6.7\% and --\,6.2\%.
\begin{table}[htp]\centering
\caption{Summary of systematic errors pertaining to the \tn\ measurement of \zr .} \begin{tabular}{lcc} \\
\toprule
Description                                  & Syst. Error (\%) & \\
\midrule
the tracker and reconstruction               & \plm 5.0 & \cite{Arnold:2004xq} \\
$\pm 1 \%$ energy calibration precision      & +\,2.9, --\,2.2 & \\
the mass of $^{96}$Zr                        & \plm 2.0 & \cite{Arnold:2004xq} \\
$\pm 10 \%$ external background precision    & \plm 0.3 & \cite{Argyriades:2009vq} \\
$\pm 10 \%$ \nd\ precision                   & \plm 0.7 & \cite{Argyriades:2008pr} \\
$\pm 5 \%$ internal background precision     & \plm 1.9 & \\
the likelihood fit energy window             & +\,1.6, --\,0.2 & \\
\midrule
\textbf{Total Systematic Error}              & \textbf{+\,6.7\%, --\,6.2\%} & \\
\bottomrule
\end{tabular} \label{table:syst} \end{table}
The final result for the $2\nu\beta\beta$ half-life of $^{96}$Zr including statistical and systematic errors is 
\begin{equation}
\ttn = \zrhalf \,.
\label{equ:2v-result} \end{equation}
For comparison, (\ref{equ:2v-result}) is consistent and \aprox 4 times more precise than the previous direct measurement $(2.1^{+0.8}_{-0.4}(stat) \pm 0.2(syst))\times10^{19}\;\textrm{yr}$~\cite{ Arnold:1999vg}.

\subsection{$2\nu$ NME}
The largest uncertainty in the effective Majorana mass determination is due to the uncertainty of the $0\nu\beta\beta$ NME ($M^{0\nu}$). It is still difficult to calculate the NMEs for $^{96}$Zr and currently there are no large-scale shell model calculations (see review~\cite{ Suhonen:1998ck}). The current models for the $M^{0\nu}$ computation of $^{96}$Zr are the quasi-particle random-phase approximation (QRPA) and the renormalized (RQRPA) \cite{ Kortelainen:2007mn, Simkovic:2007vu, Rodin:2007fz}, but unfortunately cannot precisely predict $M^{2\nu}$ due to strong dependence on the unknown parameter $g_{pp}$ (particle-particle coupling). In fact, extracted experimental values of $M^{2\nu}$ are needed to fix $g_{pp}$ which is used for the $M^{0\nu}$ computations. Two values for the parameter $g_{A}$ are generally agreed upon and the NME is computed using both $g_{A} = 1.0$ and $g_{A} = 1.25$. Recently a new approach (Projected Hartree-Fock-Bogoliubov -- PHFB model) was developed~\cite{ Chandra:2004he, Chaturvedi:2008zz, Chandra:2009my} which can predict the $M^{2\nu}$ and $M^{0\nu}$ values.

Using the measured value of the $^{96}$Zr $2\nu\beta\beta$ half-life~(\ref{equ:2v-result}) we extract the experimental value of the corresponding NME according to the formula
\begin{equation}
[T_{1/2}^{2\nu}]^{-1} = G^{2\nu} \vert M^{2\nu} \vert ^{2} \,,
\label{equ:2vbb} \end{equation}
where $G^{2\nu} = 1.8 \times 10^{-17}$\,yr$^{-1}$ is the known phase-space factor~\cite{Suhonen:1998ck} using $g_{A}=1.25$. The extracted NME is scaled by the electron rest mass and is 
\begin{equation}
M^{2\nu} = 0.049 \pm 0.002 \,.
\label{equ:M2v-result} \end{equation}
One can compare this (\ref{equ:M2v-result}) value with the calculated value, $M^{2\nu} = 0.058$~\cite{ Chandra:2004he}. The obtained precise value for $M^{2\nu}$ will be used to fix $g_{pp}$ parameter and improve the $M^{0\nu}$ calculations for $^{96}$Zr.

\subsection{$G_{F}$ Time Variation Hypothesis}
It has been suggested in~\cite{ Barabash:1999gr, Barabash:2002rg} that observed differences in half-lives of $\beta\beta$ isotopes obtained in geochemical experiments with samples of different age could be related to time dependence of the Fermi constant $G_{F}$. Due to the stronger dependence on the Fermi constant ($G_{F}^{4}$ rather than $G_{F}^{2}$), $\beta\beta$ decay offers a better sensitivity than single $\beta$ decay studies. The $^{96}$Zr\,--\,$^{96}$Mo transition is of particular interest since the daughter element is not a gas thus eliminating the main systematic error of the geochemical measurements. A comparison between the half-lives obtained with ancient zircon (ZrSiO$_{4}$) minerals characterizing the decay rate in the past with present day $\beta\beta$ decay rates obtained in a direct experiment like the one presented here allows the hypothesis to be probed with a high precision. 

A previous geochemical measurement carried out in 1992 with a $1.7 \times 10^{9}$\,yr old zircon yielded a $2\nu\beta\beta$ half-life of $(3.9 \pm 0.9) \times 10^{19}$\,yr~\cite{ Kawashima:1993zx}. An independent measurement was performed in 2001 with a number of zircons aged $\sim 1.8 \times 10^{9}$\,yr and a half-life of $(0.94 \pm 0.32) \times 10^{19}$\,yr was measured~\cite{ Wieser:2001ud}. The measurement presented in this paper~(\ref{equ:2v-result}) lies between the two geochemical measurements. More accurate studies of minerals of different age are needed in order to probe the $G_{F}$ time variation hypothesis with high precision.

\section{Summary}
The most precise measurement of the $2\nu\beta\beta$ decay half-life of $^{96}$Zr to date has been presented including the characteristics of the final state electrons (energy sum, individual electron energy, and angular distribution). Using this result the \tn\ nuclear matrix element of \zr\ has been experimentally determined, \mtn\ = \zrnme. In addition the most stringent constraints on \zn\ processes for the \zr\ isotope have been obtained. The high $Q_{\beta\beta}$ value, hence large phase-space of $^{96}$Zr, makes it an excellent choice for the study of $0\nu\beta\beta$ decay if the enrichment of this isotope in large quantities proves to be feasible.

\section*{Acknowledgment}
We thank the staff at the Modane Underground Laboratory for their technical assistance in running the NEMO-3 experiment and Vladimir I. Tretyak for providing the Monte Carlo event generator (DECAY0). We acknowledge support by the Grants Agencies of the Czech Republic, RFBR (Russia), STFC (UK), and NSF (USA).

\medskip%
\bibliographystyle{hunsrt}%
\bibliography{96zr_pub_v12}%
\end{document}

%% file: newcommands.tex
\newcommand{\zrhalf}{\ensuremath{[2.35\plm 0.14(stat)\plm 0.16(syst)]\pow{19}\,\mathrm{yr}}}
\newcommand{\zrnme}{\ensuremath{0.049\plm 0.002}}

\newcommand{\tab}[1]{Tab.\,\ref{#1}}

\newcommand{\kgy}{kg$\cdot$y}

\newcommand{\zn}{\ensuremath{0\nu\beta\beta}}
\newcommand{\tn}{\ensuremath{2\nu\beta\beta}}

\newcommand{\e}{\ensuremath{1e}}
\newcommand{\eg}{\ensuremath{e\gamma}}
\newcommand{\ee}{\ensuremath{ee}}
\newcommand{\ea}{\ensuremath{e\alpha}}
\newcommand{\egg}{\ensuremath{e\gamma\gamma}}
\newcommand{\eggg}{\ensuremath{e\gamma\gamma\gamma}}

\newcommand{\mbb}{\ensuremath{\langle m_{\beta\beta} \rangle}}
\newcommand{\rhc}{\ensuremath{\langle \lambda \rangle}}
\newcommand{\mnc}{\ensuremath{\langle g_{\chi^{0}} \rangle}}

\newcommand{\aprox}{\ensuremath{\sim}\,}
\newcommand{\pow}[1]{\ensuremath{\,\times\,10^{#1}}}

\newcommand{\lt}{\ensuremath{\,<\,}}

\newcommand{\plm}{\ensuremath{\,\pm}\,}

\newcommand{\x}{\ensuremath{\times}}

\newcommand{\tnt}{\ensuremath{T^{\,2\nu}_{1/2}}}

\newcommand{\ttn}{\tnt}

\newcommand{\mtn}{\ensuremath{M^{2\nu}}}

\newcommand{\prespace}{\,}
\newcommand{\ca}{\ensuremath{\prespace^{48}\mathrm{Ca}}}

\newcommand{\zr}{\ensuremath{\prespace^{96}\mathrm{Zr}}}

\newcommand{\nd}{\ensuremath{\prespace^{150}\mathrm{Nd}}}

\renewcommand{\k}{\ensuremath{\prespace^{40}\mathrm{K}}}

\newcommand{\bi}[1]{\ensuremath{\prespace^{#1}\mathrm{Bi}}}

\newcommand{\tl}[1]{\ensuremath{\prespace^{#1}\mathrm{Tl}}}

\newcommand{\hw}{17pc}
\newcommand{\ext}{pdf}
\newcommand{\eplots}[1]{./#1.\ext}
\newcommand{\egplots}[1]{./#1.\ext}
\newcommand{\eeplots}[1]{./#1.\ext}



%% file: nemo3-authors.tex

\author[LAL]{J.~Argyriades\corref{none}}
\author[IPHC]{R.~Arnold}
\author[LAL]{C.~Augier}
\author[INL]{J.~Baker}
\author[ITEP]{A.S.~Barabash}
\author[UCL]{A.~Basharina-Freshville}
\author[LAL]{M.~Bongrand}
\author[UB,CENBG]{G.~Broudin-Bay}
\author[JINR]{V.~Brudanin}
\author[INL]{A.J.~Caffrey}
\author[LPC]{A.~Chapon}
\author[UB,CENBG]{E.~Chauveau}
\author[UCL]{Z.~Daraktchieva}
\author[LPC]{D.~Durand}
\author[JINR]{V.~Egorov}
\author[UM]{N.~Fatemi-Ghomi}
\author[UCL]{R.~Flack}
\author[LPC]{B.~Guillon}
\author[UB,CENBG]{Ph.~Hubert}
\author[LAL]{S.~Jullian}
\author[UCL]{M.~Kauer\corref{cor1}} \ead{kauer@hep.ucl.ac.uk}
\author[UCL]{S.~King}
\author[JINR]{A.~Klimenko}
\author[JINR]{O.~Kochetov}
\author[ITEP]{S.I.~Konovalov}
\author[JINR]{V.~Kovalenko}
\author[LAL]{D.~Lalanne}
\author[USMBA]{T.~Lamhamdi}
\author[UTA]{K.~Lang}
\author[LPC]{Y.~Lemi\`{e}re}
\author[LPC]{C.~Longuemare}
\author[UB,CENBG]{G.~Lutter}
\author[IEAP]{F.~Mamedov}
\author[UB,CENBG]{Ch.~Marquet}
\author[IFIC]{J.~Martin-Albo}
\author[LPC]{F.~Mauger}
\author[UB,CENBG]{A.~Nachab}
\author[UM]{I.~Nasteva}
\author[JINR]{I.~Nemchenok}
\author[UB,CENBG,HUS]{C.H.~Nguyen}
\author[UAB]{F.~Nova}
\author[IFIC]{P.~Novella}
\author[Saga]{H.~Ohsumi}
\author[UTA]{R.B.~Pahlka}
\author[UB,CENBG]{F.~Perrot}
\author[UB,CENBG]{F.~Piquemal}
\author[LSCE]{J.L.~Reyss}
\author[UB,CENBG]{J.S.~Ricol}
\author[UCL]{R.~Saakyan}
\author[LAL]{X.~Sarazin}
\author[JINR]{Yu.~Shitov}
\author[LAL]{L.~Simard}
\author[FMFI]{F.~\v{S}imkovic}
\author[JINR]{A.~Smolnikov}
\author[UM]{S.~Snow}
\author[UM]{S.~S\"{o}ldner-Rembold}
\author[IEAP]{I.~\v{S}tekl}
\author[Jyv]{J.~Suhonen}
\author[MHC]{C.S.~Sutton}
\author[LAL]{G.~Szklarz}
\author[UCL]{J.~Thomas}
\author[JINR]{V.~Timkin}
\author[JINR,IPHC]{V.I.~Tretyak}
\author[ITEP]{V.~Umatov}
\author[IEAP]{L.~V\'{a}la}
\author[ITEP]{I.~Vanyushin}
\author[UCL]{V.~Vasiliev}
\author[CUP]{V.~Vorobel}
\author[JINR]{Ts.~Vylov}

\cortext[cor1]{Corresponding author}

\address{\vspace{0.1 in}(The NEMO--3 Collaboration)\vspace{0.05 in}}

\address[LAL]{LAL, Universit\'{e} Paris-Sud 11, CNRS/IN2P3, F-91405 Orsay, France}
\address[IPHC]{IPHC, Universit\'{e} de Strasbourg, CNRS/IN2P3, F-67037 Strasbourg, France}
\address[INL]{INL, Idaho National Laboratory, 83415 Idaho Falls, USA}
\address[INR]{INR, Institute of Nuclear Research, MSP 03680 Kyiv, Ukraine}
\address[ITEP]{ITEP, Institute of Theoretical and Experimental Physics, 117259 Moscow, Russia}
\address[CENBG]{CNRS/IN2P3, Centre d' Etudes Nucl\'{e}aires de Bordeaux Gradignan, UMR5797, F-33175 Gradignan, France}
\address[UB]{Universit\'{e} Bordeaux, CENBG, UMR 5797, F-33175 Gradignan, France}
\address[JINR]{JINR, Joint Institute for Nuclear Research, 141980 Dubna, Russia}             
\address[UCL]{University College London, WC1E 6BT London, United Kingdom}
\address[UM]{University of Manchester, M13 9PL Manchester, United Kingdom}
\address[USMBA]{USMBA, Universite Sidi Mohamed Ben Abdellah, 30000 Fes, Morocco}
\address[UTA]{University of Texas at Austin, 78712-0264 Austin, Texas, USA}
\address[IEAP]{IEAP, Czech Technical University in Prague,  CZ-12800 Prague, Czech Republic}
\address[LPC]{LPC, ENSICAEN, Universit\'{e} de Caen, CNRS/IN2P3, F-14032 Caen, France}
\address[IFIC]{IFIC, CSIC - Universitat de Valencia, Valencia, Spain}
\address[UAB]{Universitat Aut\`{o}noma Barcelona, Barcelona, Spain}
\address[Saga]{Saga University, Saga 840-8502, Japan}
\address[LSCE]{LSCE, CNRS, F-91190 Gif-sur-Yvette, France}
\address[FMFI]{FMFI, Commenius University, SK-842 48 Bratislava, Slovakia}
\address[Jyv]{Jyv\"{a}skyl\"{a} University,  40351 Jyv\"{a}skyl\"{a}, Finland}
\address[MHC]{MHC, Mount Holyoke College, 01075 South Hadley, Massachusetts, USA}
\address[CUP]{Charles University in Prague, Faculty of Mathematics and Physics, CZ-12116 Prague, Czech Republic}
\address[HUS]{Hanoi University of Science, Hanoi, Vietnam}